%% file: kruspe.tex
\documentclass[mypaper,7pt,twoside]{CoAst}
\usepackage{epsf,graphicx,fancyhdr}
\input{CoAst_wroclaw-proceedingslayo}

\def\rfr{\smallskip\par\noindent
        \hangindent=7truemm
        \hangafter=1}

\begin{document}
\sf

\chapterCoAst{Spectroscopy of the sdB pulsator HS\,2201$+$2610}
	     {R.~Kruspe, S.~Schuh, R.~Silvotti and I.~Traulsen}
\Authors{R.~Kruspe$^1$, S.~Schuh$^1$, R.~Silvotti$^2$ and I.~Traulsen$^1$} 
\Address{
  $^1$Institut f\"ur Astrophysik, Georg-August-Universit\"at G\"ottingen,
  Friedrich-Hund-Platz~1, 37077 G\"ottingen, Germany\\
  $^2$INAF (Istituto Nazionale di AstroFisica),
  Osservatorio Astronomico di Capodimonte,
  via Moiariello 16, 80131 Napoli, Italy
}
\noindent
\begin{abstract}
  We present time resolved echelle spectra of the planet-hosting
  subdwarf B pulsator HS\,2201$+$2610 and report on our efforts to 
  extract pulsational radial velocity measurements from this data.
\end{abstract}
\par
\Session{ \poster }
\Objects{HS\,2201$+$2610, HS\,0702$+$6043} 
\section*{HS\,2201$+$2610 -- a planet-hosting pulsating subdwarf B star}
Pulsating subdwarf B stars oscillate in short-period $p$-modes or
long-period $g$-modes. The sdB star HS\,2201$+$2610 (V\,391~Peg) is one
of the three known hybrids (Lutz et al.\ 2008), but it has become
famous for different reasons. From its $p$-modes ({\O}stensen et al.\
2001, Silvotti et al.\ 2002), a secular period change due to the
star's evolution has recently been measured over a period of 10 years.
Furthermore the O--C diagram has revealed a sinusoidal component which
is explained by the presence of a planetary-mass companion (Silvotti
et al.\ 2007). To determine the mass of the companion object, the
inclination of the orbit needs to be determined. Assuming alignment of
the orbital, rotational, and pulsational axes, it should be possible to
derive the stellar inclination from a combination of rotational
splitting in the photometric frequency spectrum and the projected
rotational velocity $v\sin{i}$ from measured rotational broadening of
spectral lines. The measured overall broadening must be corrected for
the broadening contribution due to pulsational radial velocities in
order to derive the rotational broadening. We therefore initially
concentrate on extracting pulsational radial velocities, which
requires pulsation phase resolved data.
\section*{HET HRS echelle spectra}
During two nights each in May and September 2007, spectra of
HS\,2201$+$2610 were taken with the HRS echelle spectrograph at the
9m-class Hobby-Eberly-Telescope. The spectra in September were
obtained as two series of 95 and 84\linebreak 20\,s-exposures at a
resolution of $\approx$30\,000. The resulting signal-to-noise at
5\,000\,\AA\ is slightly above 3 for individual spectra.  When
dividing the main pulsation period (349.5\,s) into ten phase bins and
summing (on average 15) individual spectra corresponding to one phase
bin, we obtain a pulsation phase resolved series of ten spectra, each
at a signal-to-noise of $\approx$ 9.
\par
A similar data set has been obtained for HS\,0702$+$6043 in
January 2008.
\section*{Preliminary results and outlook}
The series of spectra covers the hydrogen Balmer series from H$\beta$
to H$\delta$. These observed spectral lines can be cross correlated
with those of an appropriate model spectrum. Our template has been
chosen at the spectroscopic parameters $T_{\rm eff}$=30\,000\,{K} and
$\log{(g/\textrm{cm}\,\textrm{s}^{-2})}$=5.5, close to those given by
{\O}stensen et al.\ 2001.  From the cross correlation results for
H$\beta$ we do not reliably detect a sinusoidal variation with the
phase, and instead give an upper limit of $16${km\,s$^{-1}$} for the
pulsational radial velocity amplitude corresponding to the main
photometric period. Results for the other spectral lines and for
further pulsation periods are work in progress. This preliminary
result however already implies that the broadening due to the
pulsational radial velocities is small, and it looks like the same is
true for the rotational broadening. So while the average of all
spectra (including the May 2007 series) will help us to refine the
stellar parameters of HS\,2201$+$2610 in a future spectral analysis,
it remains a challenge to use them to put constraints on the orbit
solution for the planet.
\par
\acknowledgments{
  The authors thank H.~Edelmann for his help in obtaining the HET
  spectra, and U.~Heber for kindly providing grids of model spectra.
  This work has further benefitted from the advice of S.~Dreizler and
  R.~Lutz.  R.~Kruspe thanks HELAS for financially supporting the
  presentation of this poster through a conference fee waiver.
}
\par
\References{
\rfr Lutz~R., Schuh~S., Silvotti~R., et al.\ 2008, A\&A, submitted
\rfr {\O}stensen~R., Solheim~J.-E., Heber~U., et al.\ 2001, A\&A, 368, 175
\rfr Silvotti~R., Janulis~R., Schuh~S., et al.\ 2002, A\&A, 389, 180
\rfr Silvotti~R., Schuh~S., Janulis~R., et al.\ 2007, Nature, 449, 189
}
\end{document}

%% file: CoAst_wroclaw-proceedingslayo.tex
\renewcommand{\chaptermark}[1]{\markboth{#1}{}}

\newcommand{\kopf}{\small\itshape Comm. in Asteroseismology \\ Contribution to the Proceedings of the Wroclaw HELAS Workshop, 2008}

\newcommand{\Authors}[1]{\begin{center}\normalsize\bf\sf #1 \end{center}}

\renewcommand{\author}[1]{\begin{center}\normalsize\bf\sf #1 \end{center}}
\newcommand{\Address}[1]{\begin{center}\small\sf #1 \end{center}}

\newcommand{\Session}[1]{{\vspace{3mm}\small \noindent  \hspace*{3mm} Session: } #1 \normalsize}

\newcommand{\Objects}[1]{{\vspace{0mm}\small \noindent  \hspace*{3mm} Individual Objects: } \small #1 \normalsize}

	\newcommand{\poster}{\small Poster \newline}

\renewenvironment{abstract}{\section*{Abstract}\normalsize\sf}{}
\newcommand{\References}[1]{\begin{flushleft}{\large References\\}\vspace*{2mm}\small #1 \end{flushleft}}

\newcommand{\chapterCoAst}[2]{\chapter[\sf\normalsize #1\\ \footnotesize \hspace*{5mm}by #2 \sf\normalsize][]{#1\\}\rhead[\fancyplain{}{\sf\footnotesize \center{#1}}]{\fancyplain{}{\sffamily\thepage}}\lhead[\fancyplain{\kopf}{\sffamily\thepage}]{\fancyplain{\kopf}{\sf\footnotesize \center{#2}}}}

\newcommand{\figureCoAst}[5]{\begin{figure}[#4]
\centering
\includegraphics*[#5]{#1}
\caption{#2}
\label{#3}
\end{figure}}

\renewcommand{\chaptername}{}

\newcommand{\acknowledgments}[1]{\vspace*{5mm}\noindent  \textbf{Acknowledgments.} #1}